# Transition from synchronous to asynchronous superfluid phase slippage in an aperture array

Y. Sato, E. Hoskinson and R. E. Packard

*Department of Physics, University of California, Berkeley CA 94720, USA*

**We have investigated the dynamics of superfluid phase slippage in an array of apertures. The magnitude of the dissipative phase slips shows that they occur simultaneously in all the apertures when the temperature is near $T_\lambda - T \approx 10$ mK and subsequently lose their simultaneity as the temperature is lowered. We find that when periodic synchronous phase slippage occurs, the synchronicity exists from the very first phase slip, and therefore is not due to mode locking of interacting oscillators. When the system is allowed to relax freely from a given initial energy, the total number of phase slips that occur and the energy left in the system after the last phase slip depends reproducibly on the initial energy. We find the energy remaining after the final phase slip is a periodic function of the initial system energy. This dependence directly reveals the discrete and dissipative nature of the phase slips and is a powerful diagnostic for investigation of synchronicity in the array. When the array slips synchronously, this periodic energy function is a sharp sawtooth. As the temperature is lowered and the degree of synchronicity drops, the peak of this sawtooth becomes rounded, suggesting a broadening of the time interval over which the array slips. The underlying mechanism for the higher temperature synchronous behavior and the following loss of synchronicity at lower temperatures is not yet understood. We discuss the implications of our measurements and pose several questions that need to be resolved by a theory explaining the synchronous behavior in this quantum system. An understanding of the array phase slip process is essential to the optimization of superfluid 'dc-SQUID' gyroscopes and interferometers [1].**

Superfluid [4]He is described by a complex order parameter $\psi \propto e^{i\phi}$. Phase differences are proportional to superfluid velocity and vary as $d(\Delta\phi)/dt = -\Delta\mu/\hbar$. Superflow is driven by chemical potential differences, $\Delta\mu = m_4(\Delta P/\rho - s\Delta T)$, where $\Delta P$ and $\Delta T$ are differences in pressure and temperature, $\rho$ is the mass density, $s$ is the specific entropy, and $m_4$ is the [4]He atomic mass. One of the defining signatures of superfluidity is the absence of flow dissipation below some critical velocity, $v_c$. Whenever the flow through a submicron-size aperture reaches $v_c$, dissipation occurs in a discrete event wherein the quantum phase difference across the aperture drops by 2π [2].

Since superfluid velocity is proportional to phase gradient, this 2π "slip" corresponds to a discrete drop in velocity, $v_{slip} = \kappa/l_{eff}$ where $\kappa = h/m_4$ is the quantum of circulation and $l_{eff}$ is the effective hydrodynamic length of the aperture.

If one applies a constant chemical potential difference $\Delta\mu$ across an aperture, the superfluid velocity increases linearly to the critical velocity, followed by an abrupt drop (if the duration of the slip is short compared to the acceleration time) and followed again by a linear increase. The waveform of superfluid velocity $v_s(t)$ then resembles a sawtooth in which the phase slip events take place at an average rate equal to the Josephson frequency $f_j = \Delta\mu/h$. For single apertures, stochastic fluctuations in the critical velocity usually obscure the periodic nature of this process [3].

Recent work [4] has shown that in superfluid $^4$He periodic phase slip oscillations at frequency $f_j$ exist in an array of N (=4225) apertures. The oscillation amplitude near the superfluid transition temperature implies that the phase slips occur synchronously (i.e. simultaneously) among all the $N$ apertures. Josephson oscillations can be used as a phase difference sensor in superfluid gyroscopes and interferometers [1]. It is necessary to understand the origin of the synchronicity mechanism in order to optimize the design of such devices.

To investigate the nature of phase slips within the array, we have performed three kinds of experiments. In the first, we drive phase slip oscillations by applying a chemical potential difference across an aperture array and measure the phase slip oscillation amplitude down to $T_\lambda - T = 160$ mK. We find that the amplitude decreases rather dramatically as the temperature is lowered, as compared to what would be expected for synchronous behavior. In a second experiment, we excite transient Josephson oscillations lasting from one cycle to thousands of cycles. We find that the phase slip size does not change over many cycles of oscillation, indicating that when phase slips are synchronous, they are synchronous from the very first slip. In the third experiment, we give the system an initial excitation energy, allow it to decay through the dissipative phase slips, then record the amplitude of the sub-critical current oscillation (the so-called Helmholtz mode) that occurs after the last phase slip. We find that as the temperature decreases, phase slips within the array seem to occur in a less abrupt manner implying that a phase slip event is no longer a single simultaneous array-wide event but rather a collection of uncorrelated events localized to individual apertures. We present these three findings in the first part of this paper and discuss possible interpretations in the second.

Our experimental apparatus is shown in the inset of Fig. 1. Two volumes filled with superfluid $^4$He are separated by a diaphragm and an array of $N$ (=4225) apertures that are ~30 nm in diameter and spaced 3 μm apart in a 50 nm thick silicon nitride chip. A thin, flexible, metal-coated diaphragm can be pulled toward an electrode by the application of a voltage between them. A SQUID-based displacement sensor [5] is used to monitor the position of the diaphragm that serves as a microphone to determine the magnitude of the phase slip oscillation.

In our first type of measurement, we apply a DC step voltage between the diaphragm and the electrode at $t=0$. This pulls the flexible diaphragm toward the electrode creating a pressure head (and therefore a chemical potential difference) across the array. If the initial pull is large enough, the flow velocity inside the apertures reaches $v_c$ and the fluid undergoes $2\pi$ phase slips at the Josephson frequency. These dissipative events continue until there is no energy left to drive the fluid up to the critical velocity. The phase slip oscillation ends, and the system begins to oscillate about $\Delta\mu = 0$ at a different frequency – the Helmholtz frequency. The restoring force of the diaphragm, the inertia of the fluid moving in the apertures, and the heat capacity of the fluid in the inner volume determine the frequency of this resonant mode [6]. Fig. 1(b) shows a typical diaphragm displacement $x(t)$ during one of these relaxation transients. The discontinuities in fluid velocity due to phase slip events show up as sudden slope changes in $x(t)$. These can be seen in the first half of Fig. 1(c).

To determine whether or not phase slips are occurring synchronously throughout the array, we measure the peak-to-peak amplitude of the phase slip current oscillations, $I_{slip}$, and compare this number to the expected magnitude if all N apertures are locked together, $I_{slip}^N$. This expected magnitude is determined by directly measuring the current phase relation [7] $I(\phi)$ for the array during periods of sub-critical flow (i.e. the Helmholtz oscillation) where the flow is synchronous across the array. We are concerned with the strong coupling regime $T_\lambda - T \geq 10$ mK, where $I(\phi)$ is linear. The expected magnitude of the current oscillation for synchronous $2\pi$ phase slips is then,

$$I_{slip}^N = 2\pi \frac{dI(\phi)}{d\phi} \qquad 1.$$

If the phase slips are synchronous, $I_{slip} = I_{slip}^N$ and if the array loses synchronicity, $I_{slip} < I_{slip}^N$.

We determine mass currents through the array by monitoring the diaphragm position $x(t)$. The current through the array is given by,

$$I = \rho A \dot{x} \qquad 2.$$

where $\rho$ is the total fluid density, and $A$ is the diaphragm area.

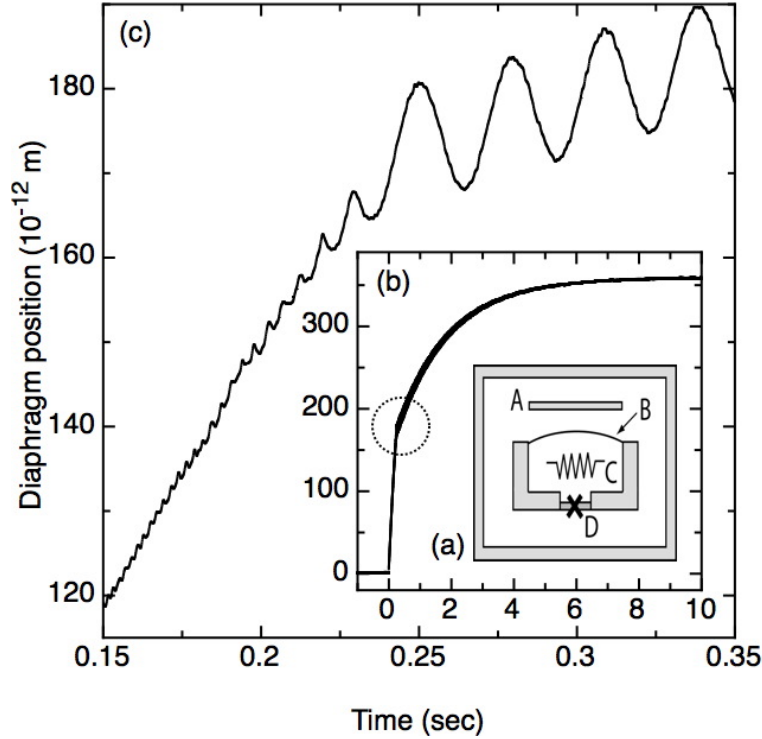

Figure 1: (a) Experimental apparatus. A: Fixed electrode. B: Soft diaphragm. C: Heater. D: Aperture array. Above the electrode is a SQUID-based transducer (not shown) which monitors the position of the diaphragm. (b) Typical diaphragm transient response. This data was taken at $T_\lambda - T \approx 9\,\text{mK}$. The pressure across the array is directly proportional to the displacement of the diaphragm from equilibrium. The initial steep rise after the pressure step at $t = 0$ is a linear relaxation during which the fluid is exhibiting phase slips at frequency $f_j$. The lightly damped Helmholtz oscillation begins when the system reaches $\Delta\mu = 0$ near $t = 0.24$ sec. The slowly curvature in the mean of the Helmholtz oscillation (between $t = 0.24$ sec and $t = 10$ sec) reflects changes in pressure head in response to a relaxing thermo-mechanical temperature differential, such that the mean $\Delta\mu$ remains zero. The dotted circle shows when the phase slip oscillation ends and the Helmholtz mode begins. The close-up is shown in (c).

When a chemical potential differential exits across the array, the diaphragm exhibits oscillations at the Josephson period, $f_j^{-1}$. If the amplitude of such diaphragm oscillations is $x_d$, the magnitude of the mass current oscillations at frequency $f_j$ is given by,

$$I_{slip} = \frac{2\pi f_j \rho A x_d}{\gamma} \qquad 3.$$

where $\gamma$ is the Fourier coefficient of the first harmonic of the displacement sensor signal. We assume here that the current exhibits a sawtooth waveform, a case where $\gamma = 2/\pi$.

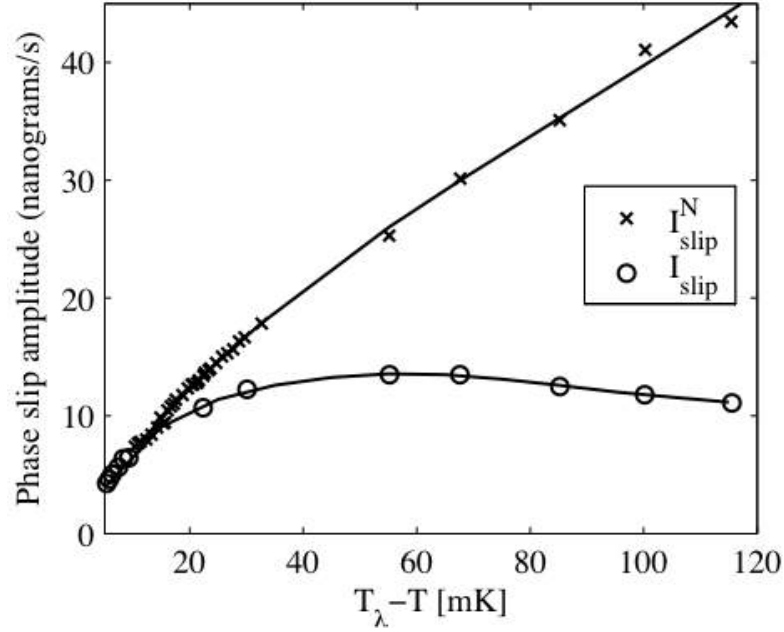

Figure 2: Measured phase slip current oscillation amplitude $I_{slip}$ (for $f_j < 300\,\text{Hz}$) and the expected value for a fully synchronous case $I_{slip}^N$. The lines are a guide to the eye.

To determine $x_d$, we record the signal $x(t)$ preceding the Helmholtz mode and compute the Fourier transform of the diaphragm oscillations. By analyzing the spectral content in small time intervals, we extract the frequency and the amplitude of the phase slip oscillations as a function of time throughout the transient. The amplitude varies with frequency due to cell resonances but levels off at lower frequencies (typically below 300Hz). We use this limiting value for $x_d$.

This Fourier analysis becomes more difficult at lower temperatures because the duration of the phase slip flow becomes shorter due to increasing critical velocity. To extend the duration of phase slip flow, we use a heater installed inside of the inner cell. First, we apply a step voltage to the heater which creates a temperature differential $\Delta T$ across the array and starts the phase slip oscillation. We then continuously increase the heater power during the transient to counteract cooling due to net superfluid flow through the array (the thermo-mechanical effect). In this way, we slow the rate at which chemical potential goes to zero. The extended transient allows us to apply the Fourier analysis described above and find the amplitude of oscillations at lower temperatures.

Once we obtain the amplitude of the diaphragm oscillations $x_d$, we use Eq. 3 to compute $I_{slip}$. Figure 2 shows the variation of $I_{slip}$ with temperature. For comparison, we also plot $I_{slip}^N$ defined by Eq. 1 using data derived from Ref.[7].

As seen in the figure, at the highest temperatures where phase slips appear, ($T_\lambda - T \approx 9$ mK), we find $I_{slip} = I_{slip}^N$, which implies that phase slips are occurring synchronously among all the $N$ apertures. However as the temperature decreases, the amplitude of current oscillation starts to rapidly decrease (relative to $I_{slip}^N$) showing a loss of synchronicity among apertures. This is the central finding of this experiment.

Neither the mechanism for the initial synchronizations nor the reason for its subsequent loss is yet understood. However, systems of interacting nonlinear oscillators often exhibit synchronization after multiple cycles [8]. If such nonlinear mode locking is present in the array, one would expect the size of first phase slip to be smaller than that of the $n^{th}$ where $n \gg 1$. Our second type of experiment is directed toward determining if there is a change in overall slip size between the first and nth phase slip oscillation.

Equation 2 shows that when a dissipative phase slip occurs, the sudden current drop in the aperture array is reflected by a sudden change in the slope of the diaphragm position curve $x(t)$. By adjusting the voltage step applied to the diaphragm we vary the length of the phase slip oscillation train from as little as one slip to as many as several thousands of slips. We then compare the abrupt slope changes, shown in Fig. 1(c), at the first slip and the $n^{th}$ slip.

The change in the slope, $\Delta \dot{x}(t)$ is determined as follows. The fluid acceleration is proportional to the chemical potential difference $\Delta \mu$ across the array. If $\Delta \mu$ is constant in the vicinity of a slip, the current increases linearly in time and the displacement of the diaphragm follows a parabola. We fit two parabolas at the cusp in the diaphragm position $x(t)$ (one before the phase slip and another right after) and find the change in the slope $\Delta \dot{x}(t)$. The result is plotted in Fig. 3. We find that the phase slip size does not change over many cycles. This result shows that when the oscillations are synchronous, they are synchronous from the very first slip. We conclude then that the synchronization is not due to a nonlinear mode locking process.

Our third experiment sheds additional light on the nature of collective phase slippage in the array. We apply a small step voltage, $V$, between the diaphragm and the electrode to create chemical potential differentials which are sufficiently small to keep the fluid velocity inside the apertures sub-critical. In the subsequent flow transient, the chemical potential reaches zero without inducing any phase slips and the diaphragm oscillates at the Helmholtz frequency with an initial amplitude $x_h$. In the absence of phase slippage, the initial energy in the Helmholtz oscillation, $E_h$, should be proportional to $E_0$, which is the energy that we put into the system by the application of a voltage step. As we increase the initial kick on the diaphragm and plot $E_h$ versus $E_0$, we expect a line with constant slope until $E_0$ is large enough to accelerate the fluid up to $v_c$, triggering a phase slip. At that point, energy is dissipated. If phase slips occur simultaneously in all the N apertures, $E_h$ should then drop discontinuously due to the abrupt extraction of energy. After such an event, as we increase $E_o$ further, $E_h$ should increase linearly again until the process repeats.

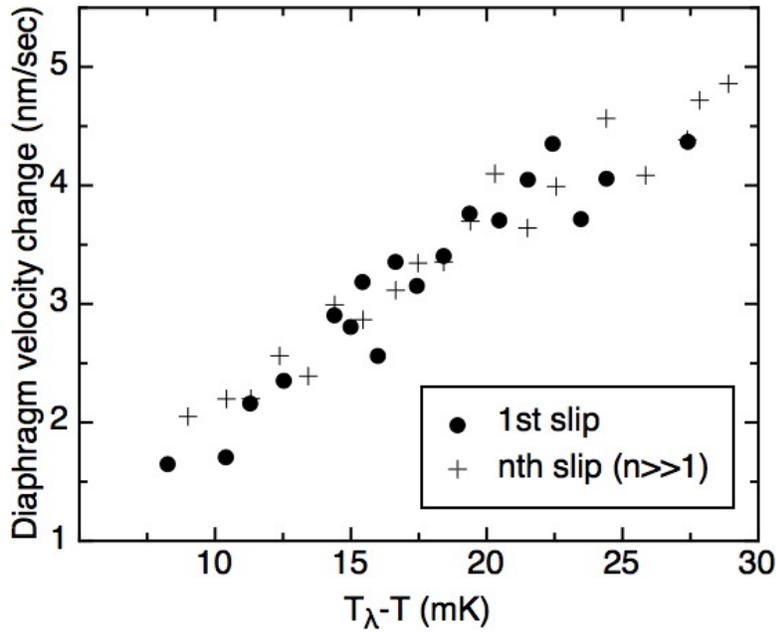

Figure 3. Diaphragm velocity change at the $1^{st}$ and the $n^{th}$ slip where $n$ is on the order of 1000. The temperature dependence comes from the increasing superfluid density as the temperature decreases.

Since the equilibrium diaphragm displacement is proportional to $V^2$, the energy that we put into the system, $E_0$, scales as $V^4$. The initial energy in the Helmholtz oscillation, $E_h$, is proportional to the square of the initial Helmholtz diaphragm oscillation amplitude, $x_h^2$. Thus a plot of $x_h^2$ versus $V^4$ (which corresponds to $E_h$ versus $E_0$) should be a sawtooth if the phase slippage occurs abruptly and simultaneously throughout the array. If the phase slippage process is distributed in time, as individual apertures slip independently of others, the sawtooth would be rounded.

Figure 4 shows our measurements of $x_h^2$ vs. $V^4$ at various temperatures. As the temperature is lowered below $T_\lambda$, the shape of $x_h^2$ vs $V^4$ evolves from a sharp sawtooth indicative of an abrupt collective phase slip event to a smoother curve that implies a continuous "phase slide" process. This suggests that some apertures are experiencing a phase slip before the others, allowing the array to dissipate energy in a more continuous manner.

Figure 4 also illustrates the striking crossover from a dissipative phase slip regime to the non-dissipative Josphson regime. The critical velocity $v_c$ (or Helmholtz amplitude) at which a slip occurs increases as the temperature decreases. At $T_\lambda - T \approx 15$ mK, $v_c \approx v_{slip}$, and a single array phase slip event removes almost all the energy in the fluid and leaves none for the Helmholtz mode. Therefore, the Helmholtz oscillation amplitude goes to 0 every time a phase slip occurs. As one gets closer to $T_\lambda$, $v_c$ becomes smaller than $v_{slip}$, and a phase slip event causes a reversal in the flow direction. Phase slips are no longer fully dissipative – the system retrieves some of the energy involved in the reversal of flow. At $T_\lambda - T \approx 5$ mK, where $v_{slip} \approx 2v_c$, dissipation due to the oscillations, which are still present as Josephson oscillations instead of phase slips, ceases. One can view this to be the complete transition into a weakly coupled Josephson regime. In the weakly coupled regime the dominant dissipation occurs through thermal conduction and normal flow – Josephson oscillations cease and Helmholtz oscillations begin when there is no longer enough energy (the flat limiting value in the 5 mK data) to reach the critical current and drive Josephson oscillations. This alternate form of dissipation, although small compared to the phase slips, explains why (in the phase slip regime) the period of the $x_h^2$ versus $V^4$ curves increases with $V^4$: for larger initial energy, the system takes longer to reach the Helmholtz mode and more energy is dissipated through thermal conduction and normal flow.

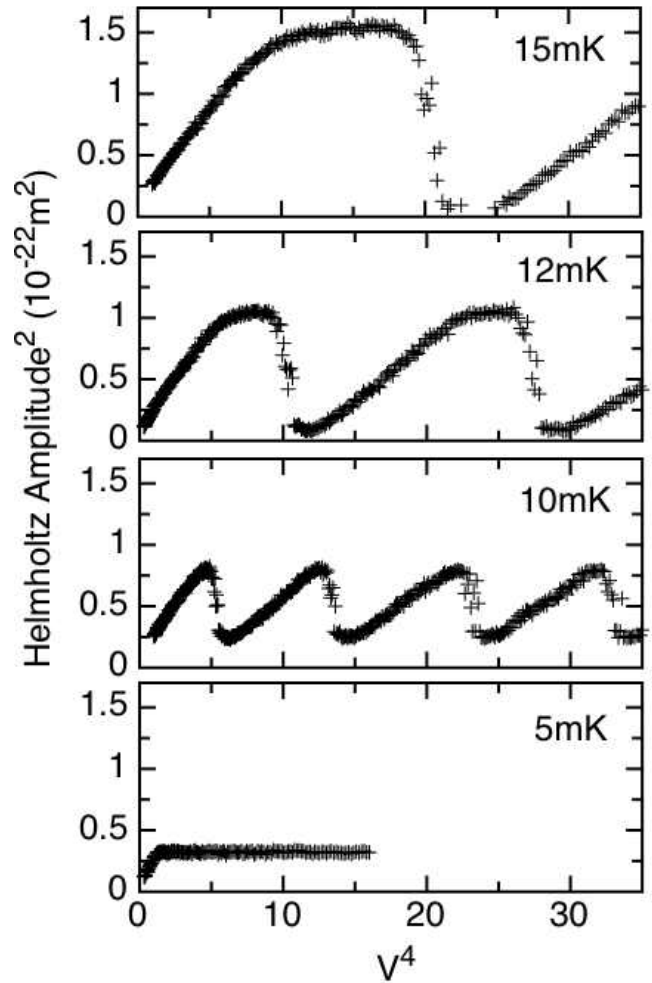

Figure 4: Measured $x_h^2$ versus $V^4$. Temperatures shown are $T_\lambda - T$ in mK.

We have considered possible mechanisms for the observed decrease in phase slip amplitude as exhibited in Fig 2. Discrete phase slippage in superfluid $^4$He is usually associated with the passage of quantized vortices that are stochastically nucleated near the aperture surface [2]. The intrinsic fluctuations cause the critical velocity to be spread out over a range $\Delta v_c$. This finite distribution width can cause the phase slip oscillation to lose its well-defined periodicity [3]. The critical velocity width $\Delta v_c$ is a function of temperature, and the relevant quantity in determining the temporal coherence of phase slip oscillations in a given aperture is $\Delta v_c/v_{slip}$. If $\Delta v_c/v_{slip} > 1$, the periodicity at $f_J$ is lost. Previous work [3, 9, 10] suggests that this ratio $\Delta v_c/v_{slip}$ increases with decreasing temperature near the superfluid transition temperature. The observed decline in the oscillation amplitude therefore could be a manifestation of loss of periodicity in any individual aperture.

Another possible mechanism for the loss of synchronicity at lower temperatures may involve variations in the surface microstructure among the array apertures. With the fluid flowing fastest near asperities, the critical velocity for an aperture must be affected by the surface inhomogeneities. Since the superfluid healing length $\xi$ is a function of temperature, how much of these nano-scale inhomogeneities the fluid actually "sees" should depend on temperature as well. The healing length is given by

$$\xi(T) = \frac{0.3 nm}{(1 - T/T_\lambda)^{0.67}} \qquad (4)$$

and it decreases from ~ 10 nm to ~ 1.5 nm as the temperature is lowered from $T_\lambda - T \approx 10$ mK to $T_\lambda - T \approx 160$ mK. If the surface variations are on the order of a few nanometers, this could very well provide a critical velocity distribution whose width increases with decreasing temperature while allowing the individual apertures to maintain well-defined periodic oscillations.

Several overarching questions remain. Is it possible for apertures to act independently in the presence of a macroscopic wavefunction? Circulation around every loop drawn through the apertures must be quantized while minimizing the energy associated with the phase gradient across the array. It is not clear how this condition is satisfied when phase slips are occurring in random positions within the array.

What are the dynamics of vortices near the transition temperature when the energy removed in a single phase slip becomes comparable to the flow energy itself? What is even meant by a "vortex" when the vortex core ~ $\xi(T)$ is comparable to the size of the apertures? Perhaps then phase slips occur by collapse of the wave function rather than by vortex dynamics [11]. The superfluid order parameter may already be so weakened that at $v_c$ the fluid in the aperture becomes momentarily normal before superfluidity is restored to a state in which the phase difference across the array has dropped by $2\pi$. This might lead to synchronicity if the wave function is so weak in all of the apertures that an excitation that causes the wave function to collapse in one aperture perturbs the other apertures enough to cause them all to collapse.

The experiments described above show that near $T_\lambda$ phase slippage occurs collectively in all the apertures in an array and the related oscillations at the Josephson frequency are not due to nonlinear synchronization. The observed decline in phase slip oscillation amplitude and the rounding of the sawtooth in the $x_h^2$ versus $V^4$ plot both indicate that array phase slippage loses its collective nature as the temperature is lowered. The results reported herein raise fundamental questions about the phase slippage process and the dynamics of a weakened superfluid confined in a multiply connected region.

We acknowledge stimulating conversations with Prof. Dung Hai Lee and Henry Fu. We thank Aditya Joshi for his suggestions to the manuscript. This work was supported in part by the NSF grant DMR 0244882 and NASA.